# α-accompanied cold ternary fission of $^{238-244}$Pu isotopes in equatorial and collinear configuration

K. P. Santhosh, Sreejith Krishnan and B. Priyanka

School of Pure and Applied Physics, Kannur University, Swami Anandatheertha Campus, Payyanur 670327, Kerala, INDIA

The cold ternary fission of $^{238}$Pu, $^{240}$Pu, $^{242}$Pu and $^{244}$Pu isotopes, with $^{4}$He as light charged particle, in equatorial and collinear configuration has been studied within the Unified ternary fission model (UTFM). The fragment combination $^{100}$Zr+$^{4}$He+$^{134}$Te possessing the near doubly magic nuclei $^{134}$Te (N=82, Z=52) gives the highest yield in the alpha accompanied ternary fission of $^{238}$Pu. For the alpha accompanied ternary fission of $^{240}$Pu, $^{242}$Pu and $^{244}$Pu isotopes, the highest yield was found for the fragment combination with doubly magic nuclei $^{132}$Sn (N=82, Z=50) as the heavier fragment. The deformation and orientation of fragments have also been taken into account for the alpha accompanied ternary fission of $^{238-244}$Pu isotopes, and it has been found that in addition to closed shell effect, ground state deformation also plays an important role in determining the isotopic yield in the ternary fission process. The emission probability and kinetic energy of long range alpha particles have been calculated and are found to be in good agreement with the experimental data.

## I. INTRODUCTION

The breakup of radioactive nuclei into more than two fission fragments has been considered as a very rare process and the formation of three fission fragments through the spontaneous fission of a radioactive nucleus is usually referred to as ternary fission. Usually, one of the ternary fission fragments is very light compared to the main fission fragments and hence the ternary fission is often referred to as light charged particle (LCP) accompanied fission. In most cases of ternary fission, the light charged particle is an alpha particle emitted in a direction perpendicular to the other two fission fragments. Such long range emission of alpha particle with an average energy of 16MeV through ternary fission was reported for the first time by Alvarez et al (see Ref. [1]), where the authors have computed the number of alpha particles emitted in coincidence with the fission of $^{235}$U and $^{239}$Pu isotopes. Later Perfilov et al. [2-4] have studied the energy spectra of alpha particles for ternary fission of different isotopes of uranium, plutonium and curium isotopes and in 1964 Malkin et al. [5] have studied the spectrum of long range alpha particle from the spontaneous fission of $^{244}$Cm isotope and obtained the most probable energy of the alpha particle as 15.8MeV. The angular distributions and the energy distributions of long range alpha particles emitted from the spontaneous ternary fission of $^{252}$Cf isotope have been studied using the statistical theory of nuclear fission by Vitta [6] and the multiparameter measurements are performed by Theobald et al., in order to compute the kinetic energies and relative angles of the three particles emitted in the ternary fission process [7]. Using triple gamma coincidence technique, the alpha accompanied ternary fission of $^{252}$Cf isotope has been studied by Ramayya et al. [8,9] and the correlated pairs Kr-Nd, Sr-Ce, Zr-Ba, Mo-Xe, Ru-Te and Pd-Sn were found to be the favorable fragment combinations.

During the last decades the ternary decay mode has been studied in detail theoretically [10-14]. Without including the pre-formation factors, a coplanar three cluster model was developed by Sandulescu et al. [15] to study the cold alpha accompanied ternary fission of $^{252}$Cf. Later in the beginning of the twenty-first century Florescu et al. [16] calculated the pre-formation amplitude for $^4$He and $^{10}$Be clusters formed in the ternary fission of $^{252}$Cf. In terms of spheroidal co-ordinates, Delion et al. [17] studied the dynamics of cold ternary fission of $^{252}$Cf isotope with light charged particle as $^4$He and $^{10}$Be. The clusters heavier than alpha particles like $^5$He, $^7$He and $^8$Li that formed during the spontaneous ternary fission of $^{252}$Cf has been studied by Kopatch et al. [18].

The studies on the emission probabilities and yield of long range alpha particle (LRA) emitted during the spontaneous fission of $^{238,240,242,244}$Pu isotopes have been measured by Serot and Wagemans [19], where the authors demonstrated that, the LRA emission strongly depended on alpha cluster preformation probability $S_\alpha$. Later, the authors [20] have also studied the long range alpha emission probability using one dimensional sudden approximation. Recently, the emission probability of long range alpha particle emitted during the spontaneous ternary fission of $^{250,252}$Cf and $^{244,246,248}$Cm isotopes was studied by Vermote et al. [21-22].

The three cluster model (TCM) developed by Manimaran et al. [23-26] has been extensively used by the authors for the ternary fission of $^{252}$Cf isotope for all fragment combinations, for the fission fragments in both the equatorial and collinear configurations. Later, Vijayaraghavan et al. [27] studied the ternary fission of $^{252}$Cf isotope for different positioning of the fragments starting from collinear configuration to the triangular configuration. Various theoretical groups [28-30] have developed several theoretical models to study the collinear cluster tri-partition, a new decay mode of heavy nuclei, for the spontaneous ternary fission of $^{252}$Cf and $^{236}$U nuclei.

Within the Unified ternary fission model (UTFM), in the present work, we are interested in the study of alpha accompanied ternary fission of $^{238-244}$Pu isotopes, with the fission fragments in equatorial and collinear configuration. The role of quadrupole deformation, emission probability of alpha particle and its kinetic energy are also studied in detail. The calculations have been done by taking the interacting barrier as the sum of Coulomb and proximity potential. Recently, using UTFM, we had evaluated the maximum yield obtained for the $^{4}$He, $^{10}$Be and $^{14}$C accompanied ternary fission of $^{242}$Cm isotope [31] and the emission probability of long range alpha particle for the various fragmentations were also calculated. Later, the emission probabilities and kinetic energies of long range alpha particles emitted from the alpha accompanied ternary fission of even-even $^{244-252}$Cm isotopes were calculated [32] and the favorable fragment combinations were singled out for both the equatorial and collinear configurations. The effect of deformation and orientation was also included and our study revealed that in addition to closed shell effect, ground state deformation plays an important role in the ternary fission of $^{244-252}$Cm isotopes. In the case of equatorial and collinear configuration, the $^{34}$Si accompanied ternary fission of $^{242}$Cm isotope has been studied and it was found that the collinear configuration is the most favorable compared to equatorial one [33]. Recently, the alpha accompanied ternary fission of $^{252}$Cf isotope has been studied with the fragments in equatorial and collinear configurations and the obtained relative yield is compared with experimental data [34]. We would like to mention that the present work is an extension of our previous works reported in Ref. [32] and this study has been performed to check the applicability of our model UTFM, on the alpha accompanied ternary fission of Pu isotopes.

The formalism used for our calculation is described in Section II. The results and discussion on the alpha accompanied ternary fission of $^{238}$Pu, $^{240}$Pu, $^{242}$Pu and $^{244}$Pu isotopes in the equatorial and collinear configuration is given in Section III and we summarize the entire work in Section IV.

## II. UNIFIED TERNARY FISSION MODEL (UTFM)

The light charged particle accompanied ternary fission is energetically possible only if $Q$ value of the reaction is positive. ie.

$$Q = M - \sum_{i=1}^{3} m_i > 0 \quad (1)$$

Here $M$ is the mass excess of the parent and $m_i$ is the mass excess of the fragments. The interacting potential barrier for a parent nucleus exhibiting cold ternary fission consists of Coulomb potential and nuclear proximity potential of Blocki et al [35, 36]. The proximity potential was first used by Shi and Swiatecki [37] in an empirical manner and has been quite extensively used by Gupta et al., [38] in the preformed cluster model (PCM) and is based on pocket formula of Blocki et al [35]. But in the present manuscript, another formulation of proximity potential (eqn (21a) and eqn (21b) of Ref. [36]) is used as given by Eqs. 7 and 8. The interacting potential barrier is given by,

$$V = \sum_{i=1}^{3} \sum_{j>i}^{3} (V_{Cij} + V_{Pij}) \quad (2)$$

with $V_{Cij} = \dfrac{Z_i Z_j e^2}{r_{ij}}$, the Coulomb interaction between the fragments. Here $Z_i$ and $Z_j$ are the atomic numbers of the fragments and $r_{ij}$ is the distance between fragment centres. The nuclear proximity potential [35] between the fragments is,

$$V_{Pij}(z) = 4\pi \gamma b \left[ \frac{C_i C_j}{(C_i + C_j)} \right] \Phi\left(\frac{z}{b}\right) \quad (3)$$

Here $\Phi$ is the universal proximity potential and $z$ is the distance between the near surfaces of the fragments. The distance between the near surfaces of the fragments for equatorial configuration is considered as $z_{12} = z_{23} = z_{13} = z$ and for collinear configuration the distance of separation are $z_{12} = z_{23} = z$ and $z_{13} = 2(C_2 + z)$. In collinear configuration the second fragment ($^4$He) is considered to lie in between the first and third fragment. The Süssmann central radii $C_i$ of the fragments related to sharp radii $R_i$ is,

$$C_i = R_i - \left(\frac{b^2}{R_i}\right) \quad (4)$$

For $R_i$ we use semi empirical formula in terms of mass number $A_i$ as [35]

$$R_i = 1.28 A_i^{1/3} - 0.76 + 0.8 A_i^{-1/3} \quad (5)$$

The nuclear surface tension coefficient called Lysekil mass formula [39] is

$$\gamma = 0.9517\,[1-1.7826\,(N-Z)^2/A^2]\ \text{MeV/fm}^2 \tag{6}$$

where N, Z and A represents neutron, proton and mass number of the parent, Φ, the universal proximity potential (eqn (21a) and eqn (21b) of Ref. [36]) is given as,

$$\Phi(\varepsilon) = -4.41 e^{-\varepsilon/0.7176}\ ,\ \text{for } \varepsilon > 1.9475 \tag{7}$$

$$\Phi(\varepsilon) = -1.7817 + 0.9270\varepsilon + 0.0169\varepsilon^2 - 0.05148\varepsilon^3\ ,\ \text{for } 0 \le \varepsilon \le 1.9475 \tag{8}$$

with $\varepsilon = z/b$, where the width (diffuseness) of the nuclear surface $b \approx 1$ fermi.

Using one-dimensional WKB approximation, barrier penetrability P, the probability which the ternary fragments cross the three body potential barrier is given as

$$P = \exp\left\{-\frac{2}{\hbar}\int_{z_1}^{z_2}\sqrt{2\mu(V-Q)}\,dz\right\} \tag{9}$$

The turning points $z_1 = 0$ represent touching configuration and $z_2$ is determined from the equation $V(z_2) = Q$, where Q is the decay energy. The potential V in eqn. 9, which is the sum of Coulomb and proximity potential given by eqn. 2, are computed by varying the distance between the near surfaces of the fragments. In eqn. 9 the mass parameter is replaced by reduced mass $\mu$ and is defined as,

$$\mu = m\left(\frac{\mu_{12} A_3}{\mu_{12} + A_3}\right) \tag{10}$$

and

$$\mu_{12} = \frac{A_1 A_2}{A_1 + A_2} \tag{11}$$

where m is the nucleon mass and $A_1$, $A_2$ and $A_3$ are the mass numbers of the three fragments. The relative yield can be calculated as the ratio between the penetration probability of a given fragmentation over the sum of penetration probabilities of all possible fragmentation as follows,

$$Y(A_i, Z_i) = \frac{P(A_i, Z_i)}{\sum P(A_i, Z_i)} \tag{12}$$

### III. RESULTS AND DISCUSSIONS

The ternary fragmentation of $^{238}$Pu, $^{240}$Pu, $^{242}$Pu and $^{244}$Pu with $^4$He as light charged particle for the equatorial and collinear configurations are studied using the concept of cold reaction valley which was introduced in relation to the structure of minima in the so called driving potential. According to the Quantum mechanical fragmentation theory (QMFT) [40], the role of shell effects (largest for a spherically closed or nearly closed shell nucleus) arises through "cold reaction" or "cold decay" valleys, corresponding to the potential energy minima in the calculated fragmentation potential. The driving potential is defined as the difference between the interaction potential $V$ and the decay energy $Q$ of the reaction. The relative yield is calculated by taking the interacting potential barrier as the sum of Coulomb and proximity potential. The $Q$ values are calculated using the recent mass tables of Wang et al. [41]. The driving potential ($V$-$Q$) for the parent nucleus is calculated (keeping third fragment $A_3$ as fixed) for all possible fragments as a function of mass and charge asymmetries respectively given as $\eta = \dfrac{A_1 - A_2}{A_1 + A_2}$ and $\eta_Z = \dfrac{Z_1 - Z_2}{Z_1 + Z_2}$, at the touching configuration. For every fixed mass pair ($A_1$, $A_2$) a pair of charges is singled out for which the driving potential is minimized. Figure 1 represents the schematic diagram for the equatorial and collinear emission of three spherical fragments at the touching configuration.

## A. ALPHA ACCOMPANIED TERNARY FISSION OF $^{238}$Pu

The driving potential is calculated for the ternary fragmentation of $^{238}$Pu, treating $^4$He as the light charged particle (LCP) and is plotted as function of fragment mass number $A_1$ as shown in figure 2. The minima occur in the cold valley is for $A_1$ = $^4$He, $^{10}$Be, $^{14}$C, $^{22}$O, $^{26}$Ne, $^{30}$Mg, $^{34}$Si, $^{40}$S, $^{46}$Ar, $^{50}$Ca, $^{52}$Ca, $^{56}$Ti, $^{60}$Cr, $^{64}$Fe, $^{68}$Ni, $^{76}$Zn, $^{82}$Ge etc. The minimum is found for the fragment configuration $^{26}$Ne+$^{208}$Pb+$^4$He, and is due to the doubly magic $^{208}$Pb (N=126, Z=82). The second minimum valley is found around $^{82}$Ge for the fragment combinations $^{76}$Zn+$^{158}$Sm+$^4$He, $^{80}$Ge+$^{154}$Nd+$^4$He, $^{82}$Ge+$^{152}$Nd+$^4$He, $^{84}$Se+$^{150}$Ce+$^4$He, $^{86}$Se+$^{148}$Ce+$^4$He and is likely to be the possible fission fragments. Another deep valley occurs around $^{130}$Sn for the fragment combinations $^{98}$Sr+$^{136}$Xe+$^4$He, $^{100}$Zr+$^{134}$Te+$^4$He, $^{102}$Mo+$^{132}$Sn+$^4$He, $^{104}$Mo+$^{130}$Sn+$^4$He, $^{106}$Mo+$^{128}$Sn+$^4$He.

The barrier penetrability is calculated for each charge minimized fragment combinations found in the alpha accompanied cold ternary fission of $^{238}$Pu using eqn. 9. The relative yield is calculated and is plotted as a function of mass numbers $A_1$ and $A_2$ as shown in figure 3(a). While analyzing the plot, it is clear that, the fragment combination $^{100}$Zr+$^{134}$Te+$^4$He with $^4$He as LCP possess highest yield due to the presence of near doubly magic nucleus $^{134}$Te (Z=52, N=82). The second highest yield is observed for the fragment combination $^{104}$Mo+$^{130}$Sn+$^4$He due to the presence of near doubly magic nucleus $^{130}$Sn (Z=50, N=80). The other fragment combinations observed in the alpha accompanied ternary fission of $^{238}$Pu nucleus are $^{108}$Mo+$^{126}$Sn+$^{4}$He, $^{106}$Mo+$^{128}$Sn+$^{4}$He, $^{102}$Mo+$^{132}$Sn+$^{4}$He, $^{96}$Sr+$^{138}$Xe+$^{4}$He. Among these, the first two reactions can be attributed to the proton shell closure at Z=50 for $^{126}$Sn and $^{128}$Sn, respectively. The fragment combination $^{102}$Mo+$^{132}$Sn+$^{4}$He is due to the presence of doubly magic nucleus $^{132}$Sn (Z=50, N=82). Also, the fragment combination $^{96}$Sr+$^{138}$Xe+$^{4}$He is observed due to the presence of near neutron closure of $^{138}$Xe at N=84.

### B. ALPHA ACCOMPANIED TERNARY FISSION OF $^{240}$Pu

The driving potential is calculated for each charge minimized fragment combinations found in the ternary fission of $^{240}$Pu with $^{4}$He as light charged particle (LCP) and is plotted as a function of fragment mass number $A_1$ as shown in figure 4. In the cold valley plot the minima is found for the fragment combination with $A_1$ = $^{4}$He, $^{8}$Be, $^{10}$Be, $^{14}$C, $^{22}$O, $^{26}$Ne, $^{28}$Ne, $^{30}$Mg, $^{34}$Si, $^{36}$Si, $^{40}$S, $^{42}$S, $^{46}$Ar, $^{50}$Ca, $^{52}$Ca, $^{56}$Ti, $^{60}$Cr etc. The deep valley occurs around $^{134}$Te, for the fragment combinations $^{102}$Zr+$^{134}$Te+$^{4}$He, $^{104}$Mo+$^{132}$Sn+$^{4}$He, $^{106}$Mo+$^{130}$Sn+$^{4}$He, $^{108}$Mo+$^{128}$Sn+$^{4}$He may be the most favorable fragment combinations. Here the minima found for $^{102}$Zr+$^{134}$Te+$^{4}$He is due to the near double magicity (Z=52, N=82) of $^{134}$Te. The fragment combination with $^{104}$Mo+$^{132}$Sn+$^{4}$He and $^{106}$Mo+$^{130}$Sn+$^{4}$He is due to the presence of the doubly magic $^{132}$Sn (Z=50, N=82) and nearly doubly magic $^{130}$Sn (Z=50, N=80) respectively.

The relative yield is calculated and plotted as a function of fragment mass number $A_1$ and $A_2$, as in the figure 3(b). From the figure, it is clear that the fragment combination $^{104}$Mo+$^{132}$Sn+$^{4}$He with $^{4}$He as LCP possess highest yield due to the presence of doubly magic $^{132}$Sn (Z=50, N=82). The next higher yield can be observed for the $^{106}$Mo+$^{130}$Sn+$^{4}$He combination and is due to the nearly doubly magic $^{130}$Sn (Z=50, N=80). The various other fragment combination observed in this α-accompanied ternary fission of parent nuclei $^{240}$Pu are $^{108}$Mo+$^{128}$Sn+$^{4}$He and $^{102}$Zr+$^{134}$Te+$^{4}$He. Of these the first one is attributed to the magic shell Z=50 of $^{128}$Sn, while the second fragment combination is due to the nearly doubly closed shell Z=52 and N=82 of $^{134}$Te.

### C. ALPHA ACCOMPANIED TERNARY FISSION OF $^{242}$Pu

The driving potential for the possible fragment combinations are calculated for the alpha accompanied ternary fragmentation of $^{242}$Pu isotope and plotted as a function of fragment mass number $A_1$, as shown in figure 5. Keeping $^{4}$He as light charged particle, the minima found in the cold valley is for $A_1$ = $^{4}$He, $^{10}$Be, $^{14}$C, $^{22}$O, $^{26}$Ne, $^{32}$Mg, $^{42}$S, $^{46}$Ar, $^{52}$Ca, $^{62}$Cr, $^{78}$Zn, $^{80}$Zn, $^{82}$Ge, $^{92}$Kr, $^{104}$Zr etc. The deep valley occurs around $^{132}$Sn for the fragment combinations

$^{104}$Zr+$^{134}$Te+$^{4}$He, $^{106}$Mo+$^{132}$Sn+$^{4}$He, $^{108}$Mo+$^{130}$Sn+$^{4}$He. In this the minimum found for the combination $^{106}$Mo+$^{132}$Sn+$^{4}$He is due to the presence of doubly magic nucleus $^{132}$Sn (N=82, Z=50). Also, the minima for the combinations $^{104}$Zr+$^{134}$Te+$^{4}$He and $^{108}$Mo+$^{130}$Sn+$^{4}$He due to the presence of near doubly magic nucleus $^{134}$Te (N=82,Z=52) and $^{130}$Sn (N=80, Z=50).

The barrier penetrability is calculated for each charge minimized fragment combinations found in the cold ternary fission of $^{242}$Pu with $^{4}$He as LCP. The relative yield is calculated and plotted as a function of mass numbers $A_1$ and $A_2$ as shown in figure 3(c). From the plot it is seen that, the fragment combination $^{106}$Mo+$^{132}$Sn+$^{4}$He possess highest yield which is due to the presence of doubly magic $^{132}$Sn (N=82, Z=50). The second highest yield is found for the splitting $^{108}$Mo+$^{130}$Sn+$^{4}$He and is due to the near doubly magic nuclei $^{130}$Sn (N=80, Z=82). The various other fragment combinations observed in this alpha accompanied ternary fission of $^{242}$Pu are $^{102}$Zr+$^{136}$Te+$^{4}$He, $^{104}$Zr+$^{134}$Te+$^{4}$He and $^{110}$Mo+$^{128}$Sn+$^{4}$He. The combination $^{104}$Zr+$^{134}$Te+$^{4}$He is due to the presence of near doubly magic $^{134}$Te (N=82, Z=52). The splitting $^{110}$Mo+$^{128}$Sn+$^{4}$He is due to the proton shell closure of $^{128}$Sn at Z=50.

### D. ALPHA ACCOMPANIED TERNARY FISSION OF $^{244}$Pu

The driving potential for $^{244}$Pu as a representative parent nucleus with $^{4}$He as light charged particle (LCP) is calculated and is plotted as a function of fragment mass number $A_1$ as shown in figure 6. The minima is found for the splitting with $A_1$ = $^{4}$He, $^{10}$Be, $^{14}$C, $^{16}$C, $^{22}$O, $^{24}$O, $^{26}$Ne, $^{32}$Mg, $^{34}$Mg, $^{36}$Si,$^{42}$S, $^{46}$Ar, $^{52}$Ca, $^{62}$Cr, $^{74}$Ni, $^{78}$Zn etc. The deepest minima for the fragment combination with $^{108}$Mo+$^{132}$Sn+$^{4}$He and $^{110}$Mo+$^{130}$Sn+$^{4}$He is due to the presence of the doubly magic $^{132}$Sn (Z=50, N=82) and near doubly magic $^{130}$Sn (Z=50, N=80).

The barrier penetrability is calculated for each charge minimized fragment combination found in the cold ternary fission of $^{244}$Pu. The relative yield is calculated and plotted as a function of fragment mass number $A_1$ and $A_2$ as shown in figure 3(d). The fragment combination with $^{108}$Mo+$^{132}$Sn+$^{4}$He possess highest yield due to the presence of doubly magic nuclei $^{132}$Sn (Z=50, N=82). The other fragment combination observed in this α-accompanied ternary fission of parent nuclei $^{244}$Pu is $^{110}$Mo+$^{130}$Sn+$^{4}$He which is due to the nearly doubly closed shell effect of $^{130}$Sn ( Z=50, N=80). The next highest yield can be observed for $^{106}$Mo+$^{134}$Te+$^{4}$He combination and is due to near double magicity (Z=52, N=82) of $^{134}$Te.The next higher yield can be observed for $^{104}$Zr+$^{136}$Te+$^{4}$He.

# E. ALPHA ACCOMPANIED TERNARY FISSION OF $^{238-244}$Pu ISOTOPES IN COLLINEAR CONFIGURATION

The alpha accompanied ternary fission of $^{238-244}$Pu isotopes has been studied with fragments in collinear configuration, in which the light charged particle $^4$He lies in between the other two fission fragments. The driving potential is calculated for all possible fragment combinations of $^{238-244}$Pu isotopes and has been plotted as a function of mass numbers $A_1$ and is shown in figure 7. From the plot it is clear that, in all cases, the least driving potential is obtained for the fragment combination with $^4$He as $A_1$. But the fragment combinations with higher $Q$ values and those with doubly or near doubly magic nuclei will be the most favorable fragment combinations, and this could be clarified through the calculation of barrier penetrability. In the ternary fission of $^{238}$Pu isotope, the fragment combinations $^{100}$Zr+$^4$He+$^{134}$Te may be the most favorable as it possess near doubly magic nuclei $^{134}$Te (Z=52, N=82) and for $^{240,242,244}$Pu isotopes, the fragment combinations with the doubly magic nuclei $^{132}$Sn (Z=50, N=82) may be the most favorable.

The barrier penetrability is calculated for the alpha accompanied ternary fission of $^{238,240,242,244}$Pu isotope and hence the relative yield is calculated. In figure 8, the relative yield is plotted as a function of mass numbers $A_1$ and $A_3$ and also the fragments with higher relative yield are labeled. For $^{238}$Pu isotope, the highest yield is obtained for the fragment combination $^{100}$Zr+$^4$He+$^{134}$Te which possess near doubly magic nuclei $^{134}$Te (Z=52, N=82). The next highest yield is obtained for the fragment combination $^{104}$Mo+$^4$He+$^{130}$Sn, which also possess near doubly magic nuclei $^{130}$Sn (Z=50, N=80).

For the alpha accompanied ternary fission of $^{240}$Pu isotope, the highest yield is obtained for the fragment combination $^{104}$Mo+$^4$He+$^{132}$Sn which possess doubly magic nuclei $^{132}$Sn (Z=50, N=82). The next highest yield is obtained for the fragment combination $^{106}$Mo+$^4$He+$^{130}$Sn, in which $^{130}$Sn (Z=50, N=80) is a near doubly magic nuclei. In the case of $^{242}$Pu isotope, the highest yield is obtained for $^{106}$Mo+$^4$He+$^{132}$Sn and the next highest yield is obtained for the fragment combination $^{108}$Mo+$^4$He+$^{130}$Sn, which possess near doubly magic nuclei $^{130}$Sn (Z=50, N=80). For the $^{244}$Pu isotope, the highest yield is obtained for the fragment combination $^{108}$Mo+$^4$He+$^{132}$Sn and the next highest yield is obtained for the splitting $^{110}$Mo+$^4$He+$^{130}$Sn, which posses doubly magic nuclei $^{132}$Sn (Z=50, N=82) and near doubly magic nuclei $^{130}$Sn (Z=50, N=80) respectively.

With the fragments in collinear configuration, the ternary fission of $^{238-244}$Pu isotope has been studied with $^4$He as light charged particle and it was found that the fragments splitting with higher *Q* value and doubly or near doubly magic nuclei plays an important role for the most favorable fragment combinations. From a comparative study with the relative yield obtained from the equatorial and collinear configuration of fragments, it could be seen that, in both equatorial and collinear configuration the highest yield is obtained for the same fragment combination. Also from figures 3 and 8, it is clear that the relative yield found for the equatorial configuration is twice as that of the collinear one. Hence we can conclude that the equatorial configuration is the most preferred configuration than the collinear configuration in the alpha accompanied ternary fission of $^{238\text{-}244}$Pu isotope. We would like to mention that, if the absolute values of yields are ignored, figure 3 and figure 8 are almost the same, this is because the alpha particle is so small compared with the main fission fragments and therefore the initial configuration of equatorial and collinear configurations are almost the same.

### F. ROLE OF DEFORMATION AND ORIENTATION OF FRAGMENTS

The effect of deformation and orientation of fragments in $^{4}$He accompanied ternary fission of $^{238\text{-}244}$Pu isotopes have been analyzed taking the Coulomb and proximity potential as the interacting barrier. The Coulomb interaction between the two deformed and oriented nuclei, which is taken from [42] and which includes higher multipole deformation [43, 44], is given as,

$$V_C = \frac{Z_1 Z_2 e^2}{r} + 3Z_1 Z_2 e^2 \sum_{\lambda, i=1,2} \frac{1}{2\lambda+1} \frac{R_{0i}^{\lambda}}{r^{\lambda+1}} Y_{\lambda}^{(0)}(\alpha_i) \left[ \beta_{\lambda i} + \frac{4}{7} \beta_{\lambda i}^2 Y_{\lambda}^{(0)}(\alpha_i) \delta_{\lambda,2} \right] \quad (13)$$

with

$$R_i(\alpha_i) = R_{0i} \left[ 1 + \sum_{\lambda} \beta_{\lambda i} Y_{\lambda}^{(0)}(\alpha_i) \right] \quad (14)$$

where $R_{0i} = 1.28 A_i^{1/3} - 0.76 + 0.8 A_i^{-1/3}$. Here $\alpha_i$ is the angle between the radius vector and symmetry axis of the $i^{th}$ nuclei (see Fig.1 of Ref [43]) and it is to be noted that the quadrupole interaction term proportional to $\beta_{21}\beta_{22}$, is neglected because of its short range character.

In proximity potential, $V_P(z) = 4\pi\gamma b \overline{R} \Phi(\varepsilon)$, the deformation comes only in the mean curvature radius. For spherical nuclei, mean curvature radius is defined as $\overline{R} = \frac{C_1 C_2}{C_1 + C_2}$, where $C_1$ and $C_2$ are Süssmann central radii of fragments. The mean curvature radius, $\overline{R}$ for two deformed nuclei lying in the same plane can be obtained by the relation,

$$\frac{1}{\overline{R}^2} = \frac{1}{R_{11}R_{12}} + \frac{1}{R_{21}R_{22}} + \frac{1}{R_{11}R_{22}} + \frac{1}{R_{21}R_{12}} \tag{15}$$

The four principal radii of curvature $R_{11}$, $R_{12}$, $R_{21}$ and $R_{22}$ are given by Baltz and Bayman [45].

Figures 2, 4, 5, 6 represent the cold valley plots, the plot connecting the driving potential (*V-Q*) and the mass number $A_1$ for even-even $^{238}$Pu to $^{244}$Pu isotopes. In these plots three cases are considered (1) three fragments taken as spherical (2) two fragments ($A_1$ and $A_2$) as deformed with $0^0 - 0^0$ orientation and (3) two fragments ($A_1$ and $A_2$) as deformed with $90^0 - 90^0$ orientation. For computing driving potential we have used experimental quadrupole deformation ($\beta_2$) values taken from Ref. [46] and for those cases where the experimental values were unavailable, we have taken them from Moller et al [47]. It can be seen from these plots that in most of the cases, $0^0 - 0^0$ orientation have a low value for driving potential, but in few cases, $90^0 - 90^0$ orientation has the low value. In the former case, either both the fragments are prolate or one fragment is prolate and the other one is spherical; and in latter case both fragments are either oblate or one fragment is oblate and the other one is spherical. It can be seen that when deformation are included, the optimum fragment combination are also found to be changed. For example in the case of $^{238}$Pu isotope, the fragment combinations $^{104}$Mo+$^{4}$He+$^{130}$Sn and $^{100}$Zr+$^{4}$He+$^{134}$Te changed to $^{104}$Tc+$^{4}$He+$^{130}$In and $^{100}$Nb+$^{4}$He+$^{134}$Sb respectively. In the case of $^{240}$Pu isotope, the fragment combinations $^{106}$Mo+$^{4}$He+$^{130}$Sn and $^{102}$Zr+$^{4}$He+$^{134}$Te changed to $^{106}$Nb+$^{4}$He+$^{130}$Sb and $^{102}$Mo+$^{4}$He+$^{134}$Sn respectively. In the case of $^{242}$Pu isotope, the fragment combinations $^{100}$Sr+$^{4}$He+$^{138}$Xe and $^{106}$Mo+$^{4}$He+$^{132}$Sn changed to $^{102}$Kr+$^{4}$He+$^{138}$Ba and $^{106}$Tc+$^{4}$He+$^{132}$In respectively. In the case of $^{244}$Pu isotope, the fragment combinations $^{108}$Mo+$^{4}$He+$^{132}$Sn and $^{110}$Mo+$^{4}$He+$^{130}$Sn changed to $^{108}$Nb+$^{4}$He+$^{132}$Sb and $^{110}$Ru+$^{4}$He+$^{130}$Cd respectively.

By including the quadrupole deformation, the barrier penetrability is calculated for all possible fragment combinations that occur in the cold valley plot which have the minimum (*V-Q*) value. The computations are done using the deformed Coulomb potential and deformed nuclear proximity potential. The inclusion of quadrupole deformation ($\beta_2$) reduces the height and width of the barrier and as a result, the barrier penetrability is found to increase. By comparing the figures 9(a)-9(d) with corresponding plots for the spherical case figures 3(a)-3(d), it can be seen that fragments with highest yield are also found to be changed. For the alpha accompanied ternary fission of $^{238}$Pu and $^{240}$Pu isotope, the highest yield is found for the fragment combination $^{94}$Sr+$^{4}$He+$^{140}$Xe and $^{112}$Ru+$^{4}$He+$^{124}$Cd respectively, with the inclusion of deformation. In the case of $^{242}$Pu and $^{244}$Pu isotopes, the highest yield is found for the fragment combination $^{110}$Ru+$^{4}$He+$^{128}$Cd and $^{110}$Ru+$^{4}$He+$^{130}$Cd respectively.

For a better comparison of the result, a histogram is plotted with yield as a function of mass numbers $A_1$ and $A_2$ for the ternary fragmentation of $^{238-244}$Pu isotopes with the inclusion of quadrupole deformation $β_2$ as shown in figures 10-13. The studies on the influence of deformation in the alpha accompanied ternary fission of $^{238-244}$Pu isotopes reveal that the ground state deformation has an important role in determining the isotopic yield in the ternary fission as that of shell effect.

## G. EMISSION PROBABILITY OF LONG RANGE ALPHA PARTICLE

The alpha particles emitted during the ternary fission process possesses an average energy of around 16MeV and such alpha particles are usually referred to as long range alpha particles (LRA). The emission probability of long range alpha particle LRA is determined with the number of fission events B and hence denoted as LRA/B. According to Carjan's model [48], the emission of long range alpha particle is possible only if two conditions are satisfied. First of all, the alpha cluster must be preformed within the fissioning nucleus and secondly it must gain enough energy to overcome the Coulomb barrier of the scission nucleus. Serot and Wagemans [19] demonstrated that, the parameters like fissility parameter $Z^2/A$, the alpha cluster preformation factor $S_α$ and fission modes also plays an important role in the emission probability of long range alpha particles.

The emission probability of long range alpha particle can be written as follows:

$$\frac{LRA}{B} = S_α P_{LRA} \tag{16}$$

where $S_α$ is the spectroscopic factor or alpha cluster preformation factor and $P_{LRA}$ is the probability of alpha particle preformed in the fission nucleus.

The spectroscopic factor $S_α$ can be calculated in a semi-empirical way proposed by Blendowske et al [49] as, $S_α = b λ_e / λ_{WKB}$, where $b$ is the branching ratio for the ground state to ground state transition, $λ_e$ is the experimental α decay constant and $λ_{WKB}$ is the α decay constant calculated from the WKB approximation.

The preformation probability of alpha particle $P_{LRA}$ can be calculated as,

$$P_{LRA} = \exp\left\{-\frac{2}{\hbar}\int_{z_0}^{z_1}\sqrt{2\mu(V-Q)}dz\right\} \tag{17}$$

Here the first turning point is determined from the equation $V(z_0) = Q$, where $Q$ is the decay energy, and the second turning point $z_1 = 0$ represent the touching configuration. For the internal (overlap) region, the potential is taken as a simple power law interpolation.

Using the formalism described above, we have computed the emission probabilities of long range alpha particle in the case of $^{238}$Pu, $^{240}$Pu, $^{242}$Pu and $^{244}$Pu isotopes and the obtained results are found to be in good agreement with the experimental data [19]. The spectroscopic factors and corresponding emission probabilities of $^{238-244}$Pu isotopes are listed in Table I.

## H. KINETIC ENERGIES OF LONG RANGE ALPHA PARTICLE IN THE TERNARY FISSION OF $^{238-244}$Pu ISOTOPES

The kinetic energy of long range alpha particle emitted in the ternary fission of $^{238-244}$Pu isotopes is computed using the formalism reported by Fraenkel [50]. The conservation of total momentum in the direction of light particle and in a direction perpendicular to light particle leads to the relations,

$$(m_L E_L)^{1/2} = (m_H E_H)^{1/2}\cos\theta_R - (m_\alpha E_\alpha)^{1/2}\cos\theta_L \tag{18}$$

$$(m_H E_H)^{1/2}\sin\theta_R = (m_\alpha E_\alpha)^{1/2}\sin\theta_L \tag{19}$$

Here $m_L$, $m_H$ and $m_\alpha$ are the masses of the light, heavy and $\alpha$ particle respectively. $E_L$, $E_H$ and $E_\alpha$ represent the final energies of the light, heavy and $\alpha$ particle respectively. The kinetic energy of the long range alpha particle can be derived from eqn. (18) and eqn. (19) and is given as,

$$E_\alpha = E_L\left(\frac{m_L}{m_\alpha}\right)(\sin\theta_L\cot\theta_R - \cos\theta_L)^{-2} \tag{20}$$

Here $\theta_L$ is the angle between the alpha particle and the light particle and $\theta_R$ is the recoil angle. Fraenkel [50] suggested that the mean kinetic energy of alpha particle $E_\alpha$ = 16MeV and total fragment kinetic energy of 168MeV is obtained for the maximum value of recoil angle $\theta_R$ = 4.5º and this maximum recoil angle is obtained for $\theta_L$ = 92.25º. It is because of the fact that, in the present work we have taken the recoil angle as $\theta_R$ = 4.5º and $\theta_L$ = 92.25º.

The kinetic energy of light fragment $E_L$ can be calculated as,

$$E_L = \frac{A_H}{A_L + A_H}TKE \tag{21}$$

where the total kinetic energies of fission fragments *TKE* can be computed using the expressions taken from Herbach et al. [51] as follows,

$$TKE = \frac{0.2904 (Z_L + Z_H)^2}{A_L^{1/3} + A_H^{1/3} - (A_L + A_H)^{1/3}} \frac{A_L A_H}{(A_L + A_H)^2} \quad (22)$$

Here $A_L$ and $A_H$ are the mass numbers of light and heavy fragments respectively.

Using the formalism described above, we have calculated the kinetic energy of the long range alpha particle in the ternary fission of $^{238-244}$Pu and is given in Table II. It is to be noted that the obtained results are found to be in good agreement with the experimental data [19].

We have included the deformation effects in our present calculations and we have found that, even though the inclusion of deformation increases the emission probability, there is no appreciable change in its order. While considering the deformation effects on the kinetic energies of LRA, it was seen that, there is no appreciable change in the kinetic energies.

## IV. SUMMARY

For alpha accompanied ternary fission of even-even $^{238-244}$Pu isotopes, the relative yield is calculated by taking the interacting barrier as the sum of Coulomb and proximity potential, with fragments in equatorial and collinear configuration, within the Unified ternary fission model (UTFM). In the ternary fission of $^{238}$Pu isotopes with $^4$He as light charged particle, the highest yield is obtained for the fragment combination $^{100}$Zr+$^4$He+$^{134}$Te, which possess near doubly magic nuclei $^{134}$Te (N=82, Z=52). For the alpha accompanied ternary fission of $^{240}$Pu, $^{242}$Pu and $^{244}$Pu isotopes, the highest yield is obtained for the fragment combination $^{104}$Mo+$^4$He+$^{132}$Sn, $^{106}$Mo+$^4$He+$^{132}$Sn and $^{108}$Mo+$^4$He+$^{132}$Sn respectively, of which $^{132}$Sn (N=82, Z=50) is a doubly magic nuclei. The effect of deformation and orientation is also studied in detail and found that ground state deformation also plays an important role as that of shell effect in determining the isotopic yield in the alpha accompanied ternary fission of $^{238-244}$Pu isotopes. The kinetic energy and emission probability of alpha particle is calculated and are found to be in good agreement with the experimental data.

## ACKNOWLEDGMENTS

The author KPS would like to thank the University Grants Commission, Govt. of India for the financial support under Major Research Project. No.42-760/2013 (SR) dated 22-03-2013.

----------------------------------------------------------------------------

TABLE I. The calculated emission probability of alpha particle in the ternary fission of different Pu isotopes and the corresponding experimental data [19] are listed. The computed spectroscopic factor $S_\alpha$ and $P_{LRA}$ are also listed.

| Isotope | $S_\alpha$ | $P_{LRA}$ | $\frac{LRA}{B}$ | $\left(\frac{LRA}{B}\right)_{EXP.}$ |
|---|---|---|---|---|
| $^{238}$Pu | 0.0317 | 0.1080 | 3.42 x $10^{-3}$ | (2.76 ± 0.13) x $10^{-3}$ |
| $^{240}$Pu | 0.0421 | 0.1067 | 4.49 x $10^{-3}$ | (2.50 ± 0.14) x $10^{-3}$ |
| $^{242}$Pu | 0.0426 | 0.1335 | 5.69 x $10^{-3}$ | (2.17 ± 0.07) x $10^{-3}$ |
| $^{244}$Pu | 0.0378 | 0.1507 | 5.70 x $10^{-3}$ | (1.17 ± 0.09) x $10^{-3}$ |

TABLE II. The calculated kinetic energy of alpha particle $E_\alpha$ emitted in the ternary fragmentation of $^{238-244}$Pu isotopes and the corresponding experimental data [19].

| Fragmentation channel | $E_\alpha$ (MeV) Calc. | $E_\alpha$ (MeV) Expt. | Fragmentation channel | $E_\alpha$ (MeV) Calc. | $E_\alpha$ (MeV) Expt. |
|---|---|---|---|---|---|
| $^{238}$Pu → $^{98}$Sr + $^4$He + $^{136}$Xe | 14.62 | | $^{242}$Pu → $^{100}$Sr + $^4$He + $^{138}$Xe | 14.81 | |
| $^{238}$Pu → $^{100}$Zr + $^4$He + $^{134}$Te | 14.76 | | $^{242}$Pu → $^{102}$Zr + $^4$He + $^{136}$Te | 14.94 | |
| $^{238}$Pu → $^{102}$Mo + $^4$He + $^{132}$Sn | 14.88 | | $^{242}$Pu → $^{104}$Zr + $^4$He + $^{134}$Te | 15.06 | |
| $^{238}$Pu → $^{104}$Mo + $^4$He + $^{130}$Sn | 14.98 | | $^{242}$Pu → $^{106}$Mo + $^4$He + $^{132}$Sn | 15.16 | |
| $^{238}$Pu → $^{106}$Mo + $^4$He + $^{128}$Sn | 15.08 | 15.91 ± 0.22 | $^{242}$Pu → $^{108}$Mo + $^4$He + $^{130}$Sn | 15.25 | 15.79 ± 0.21 |
| $^{238}$Pu → $^{108}$Mo + $^4$He + $^{126}$Sn | 15.15 | | $^{242}$Pu → $^{110}$Mo + $^4$He + $^{128}$Sn | 15.33 | |
| $^{238}$Pu → $^{110}$Ru + $^4$He + $^{124}$Cd | 15.21 | | $^{242}$Pu → $^{112}$Ru + $^4$He + $^{126}$Cd | 15.39 | |
| $^{238}$Pu → $^{112}$Ru + $^4$He + $^{122}$Cd | 15.25 | | $^{242}$Pu → $^{114}$Ru + $^4$He + $^{124}$Cd | 15.43 | |
| $^{238}$Pu → $^{114}$Ru + $^4$He + $^{120}$Cd | 15.29 | | $^{242}$Pu → $^{116}$Ru + $^4$He + $^{122}$Cd | 15.46 | |
| $^{238}$Pu → $^{116}$Pd + $^4$He + $^{118}$Pd | 15.30 | | $^{242}$Pu → $^{118}$Pd + $^4$He + $^{120}$Pd | 15.48 | |
| | | | | | |
| $^{240}$Pu → $^{100}$Zr + $^4$He + $^{136}$Te | 14.79 | | $^{244}$Pu → $^{102}$Zr + $^4$He + $^{138}$Te | 14.98 | |
| $^{240}$Pu → $^{102}$Zr + $^4$He + $^{134}$Te | 14.91 | | $^{244}$Pu → $^{104}$Zr + $^4$He + $^{136}$Te | 15.10 | |
| $^{240}$Pu → $^{104}$Mo + $^4$He + $^{132}$Sn | 15.02 | | $^{244}$Pu → $^{106}$Zr + $^4$He + $^{134}$Te | 15.21 | |
| $^{240}$Pu → $^{106}$Mo + $^4$He + $^{130}$Sn | 15.12 | | $^{244}$Pu → $^{108}$Mo + $^4$He + $^{132}$Sn | 15.30 | |
| $^{240}$Pu → $^{108}$Mo + $^4$He + $^{128}$Sn | 15.20 | 16.55 ± 0.27 | $^{244}$Pu → $^{110}$Mo + $^4$He + $^{130}$Sn | 15.38 | 16.04 ± 0.25 |
| $^{240}$Pu → $^{110}$Ru + $^4$He + $^{126}$Cd | 15.27 | | $^{244}$Pu → $^{112}$Ru + $^4$He + $^{128}$Cd | 15.45 | |
| $^{240}$Pu → $^{112}$Ru + $^4$He + $^{124}$Cd | 15.32 | | $^{244}$Pu → $^{114}$Ru + $^4$He + $^{126}$Cd | 15.50 | |
| $^{240}$Pu → $^{114}$Ru + $^4$He + $^{122}$Cd | 15.36 | | $^{244}$Pu → $^{116}$Ru + $^4$He + $^{124}$Cd | 15.54 | |
| $^{240}$Pu → $^{116}$Pd + $^4$He + $^{120}$Pd | 15.38 | | $^{244}$Pu → $^{118}$Pd + $^4$He + $^{122}$Pd | 15.56 | |
| $^{240}$Pu → $^{118}$Pd + $^4$He + $^{118}$Pd | 15.39 | | $^{244}$Pu → $^{120}$Pd + $^4$He + $^{120}$Pd | 15.56 | |

(a) Equatorial

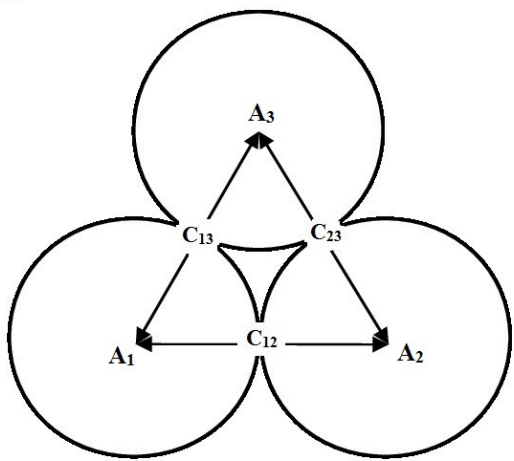

(b) Collinear

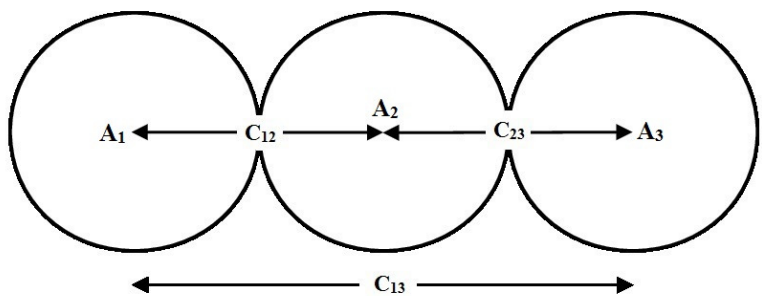

FIG 1. The touching configuration of three spherical fragments in a) equatorial configuration b) collinear configuration.

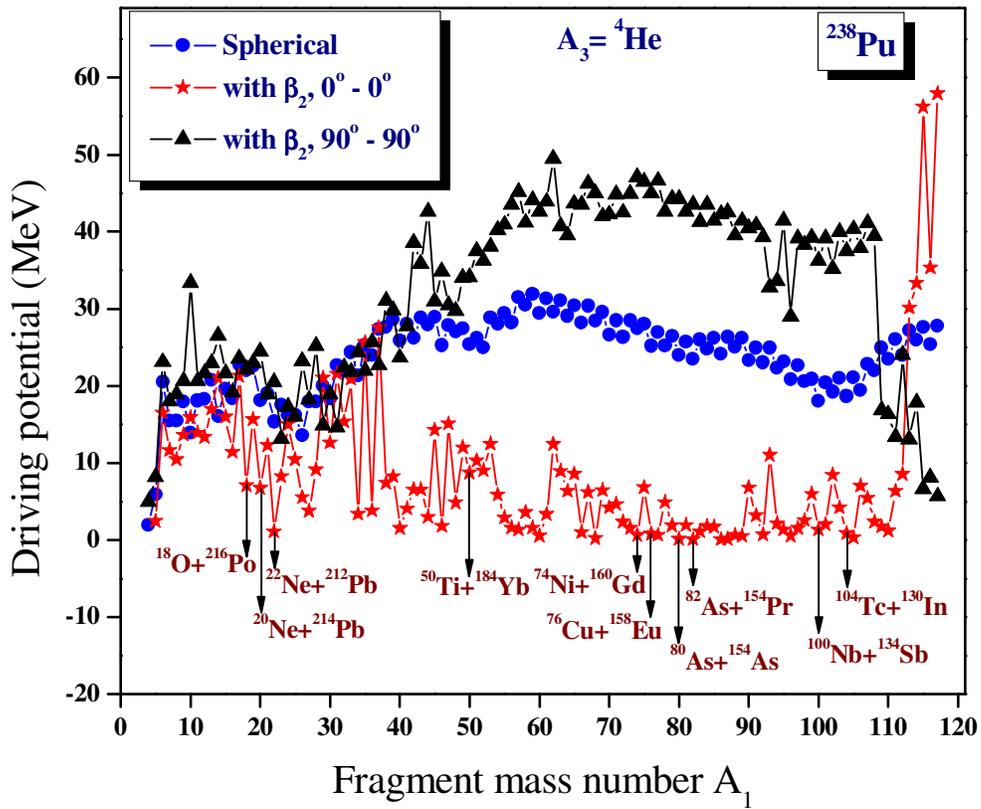

FIG 2. (Color online) The driving potential for $^{238}$Pu isotope with $^4$He as light charged particle with the inclusion of quadrupole deformation $\beta_2$ and for different orientation plotted as a function of mass number $A_1$.

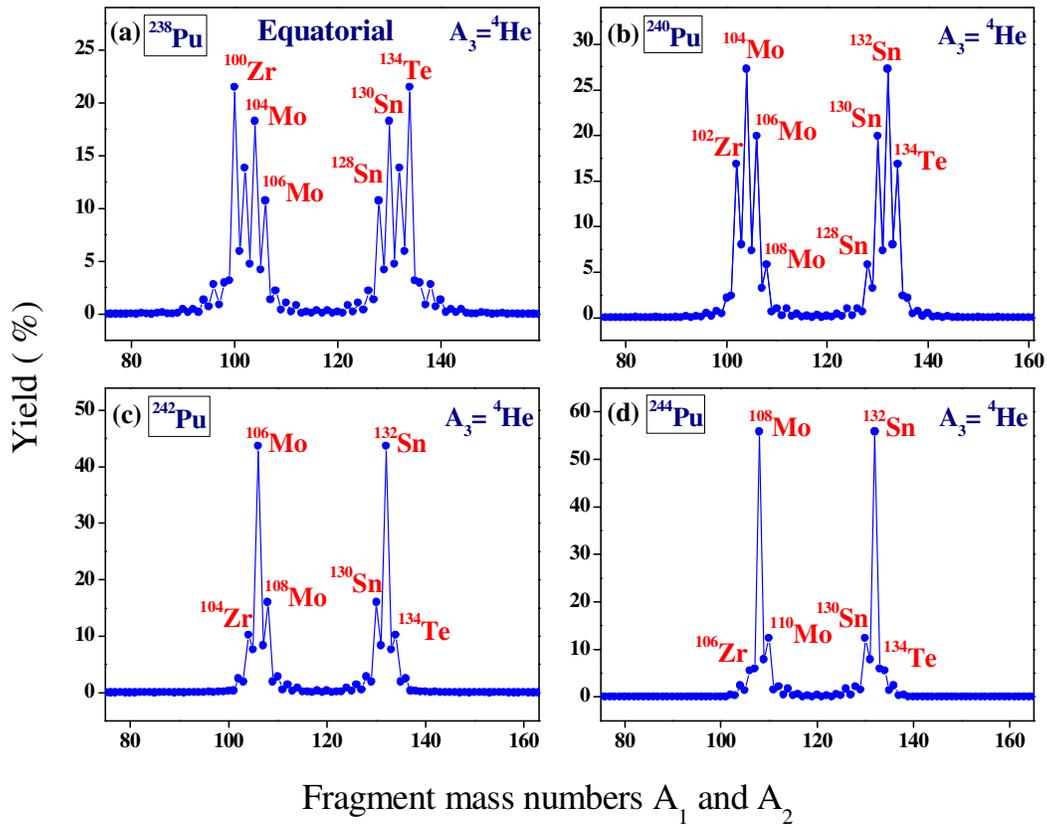

FIG 3. (Color online) The calculated yield for the charge minimized third fragment $^4$He is plotted as a function of fragment mass numbers $A_1$ and $A_2$ for the ternary fission of $^{238-244}$Pu isotopes.

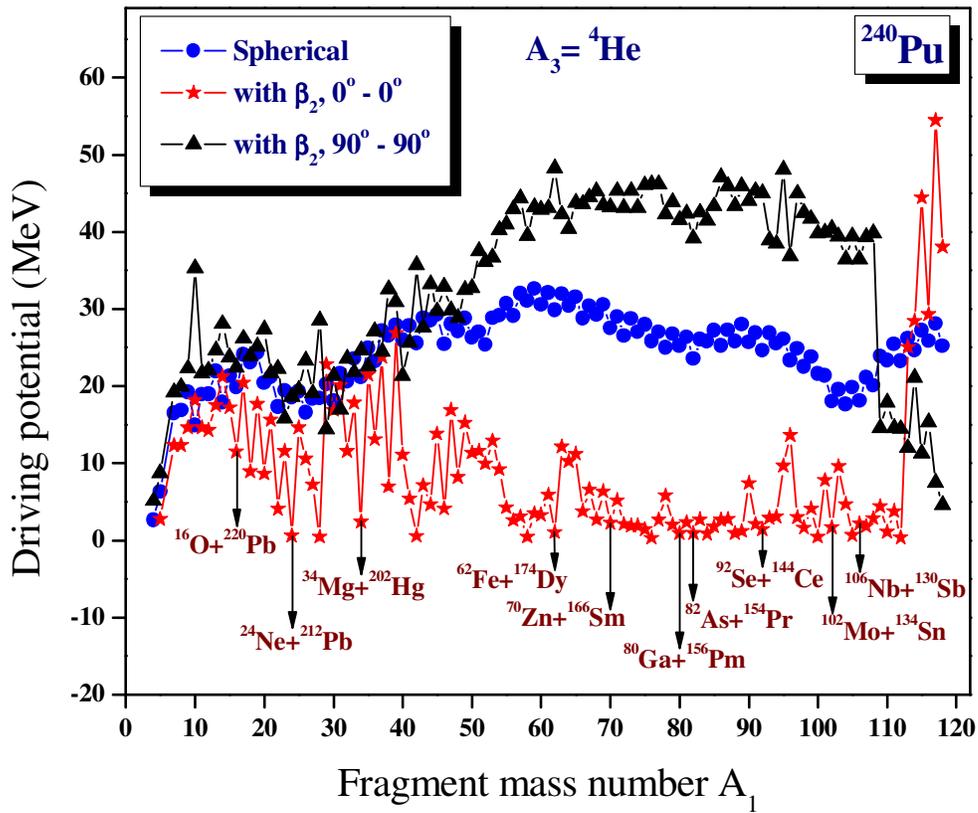

FIG 4. (Color online) The driving potential for $^{240}$Pu isotope with $^{4}$He as light charged particle with the inclusion of quadrupole deformation $\beta_2$ and for different orientation plotted as a function of mass number $A_1$.

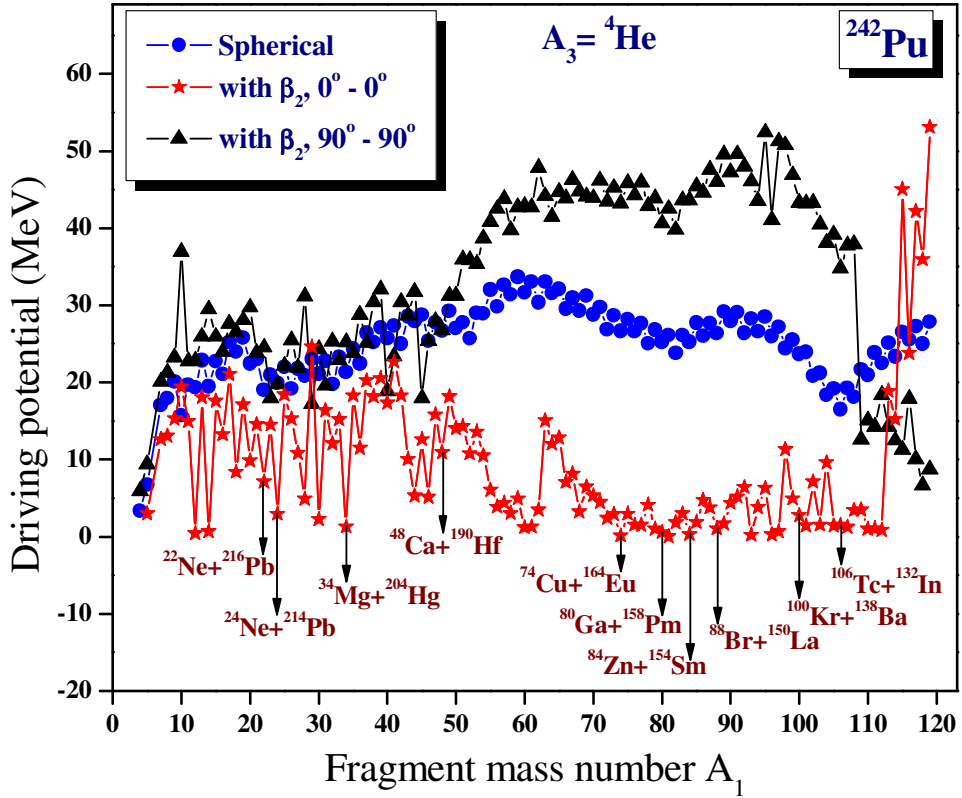

FIG 5. (Color online) The driving potential for $^{242}$Pu isotope with $^4$He as light charged particle with the inclusion of quadrupole deformation $\beta_2$ and for different orientation plotted as a function of mass number $A_1$.

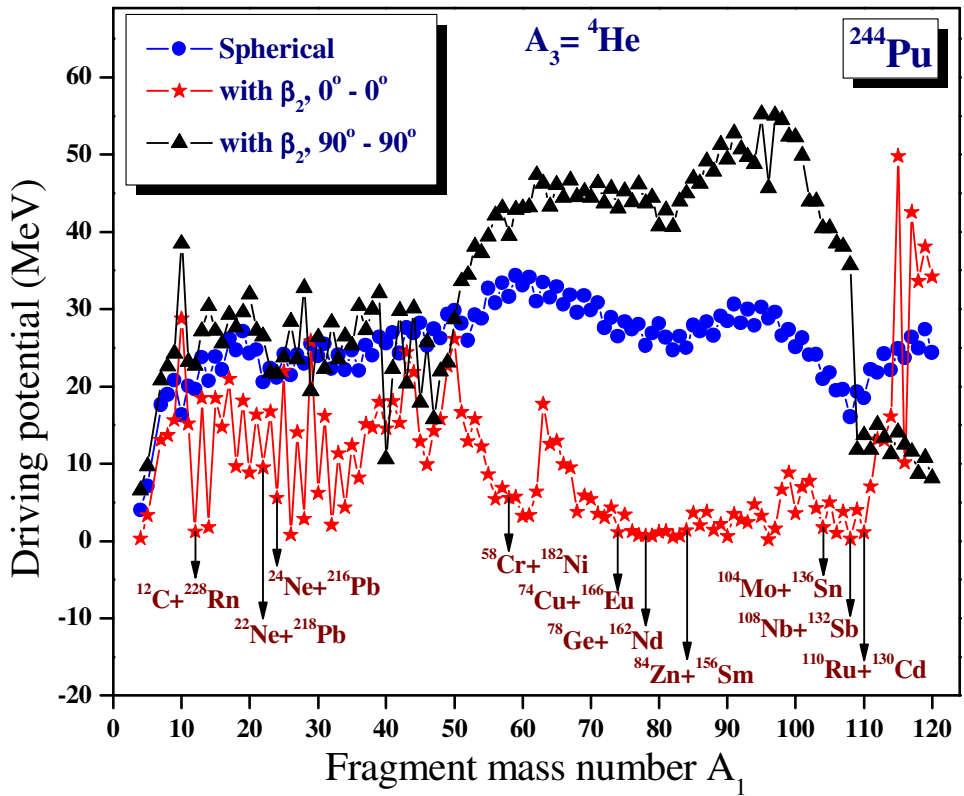

FIG 6. (Color online) The driving potential for $^{244}$Pu isotope with $^4$He as light charged particle with the inclusion of quadrupole deformation $\beta_2$ and for different orientation plotted as a function of mass number $A_1$.

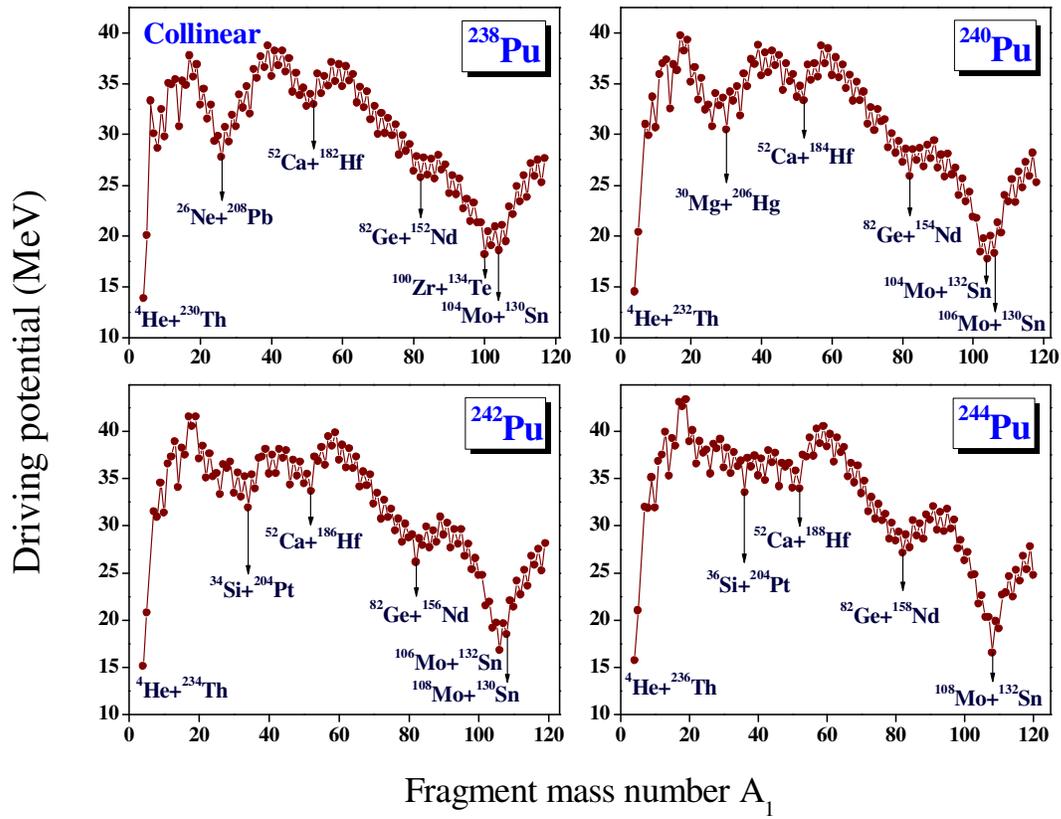

FIG 7. (Color online) The driving potential for $^{238-244}$Pu isotope with $^4$He as light charged particle in the case of collinear configuration plotted as a function of mass number $A_1$.

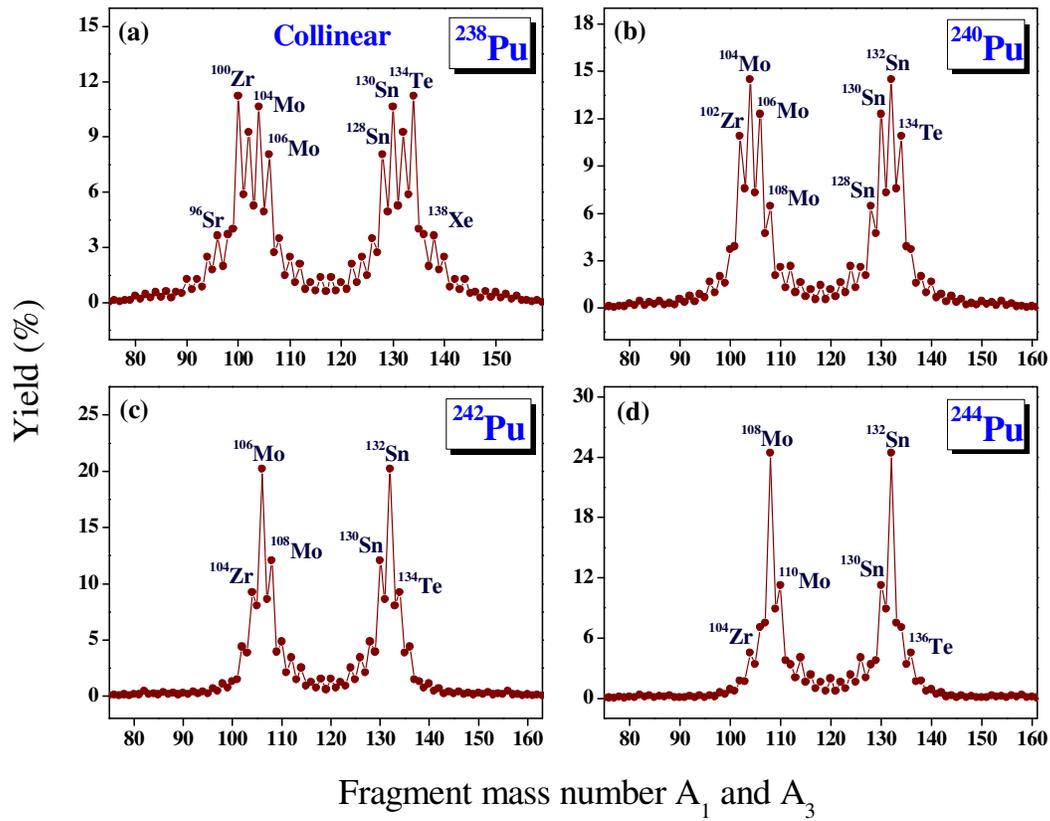

FIG 8. (Color online) The calculated yields for the ternary fission of $^{238-244}$Pu isotopes with charge minimized second fragment $^{4}$He in the case of collinear configuration plotted as a function of mass numbers $A_1$ and $A_3$. The fragment combinations with highest yield are labeled.

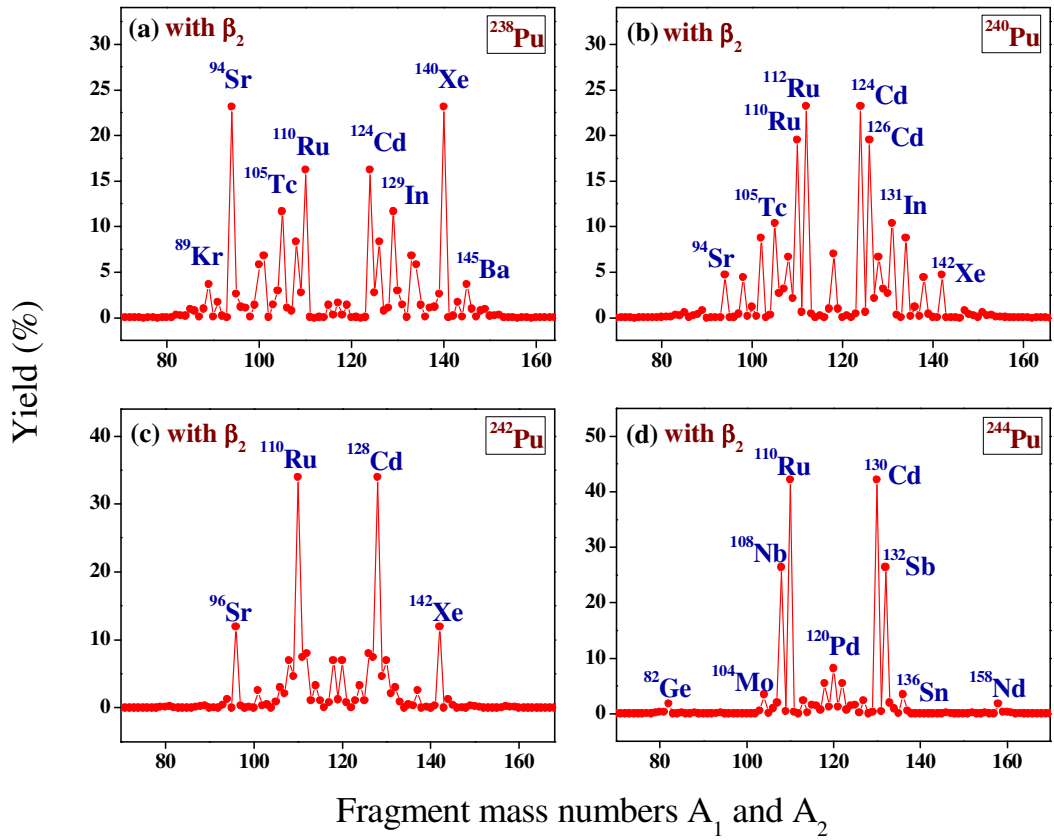

FIG 9. (Color online) The calculated yields for the charge minimized third fragment $^4$He with the inclusion of quadrupole deformation $\beta_2$ plotted as a function of mass numbers $A_1$ and $A_2$ for $^{238-244}$Pu isotopes. The fragment combinations with higher yields are labeled.

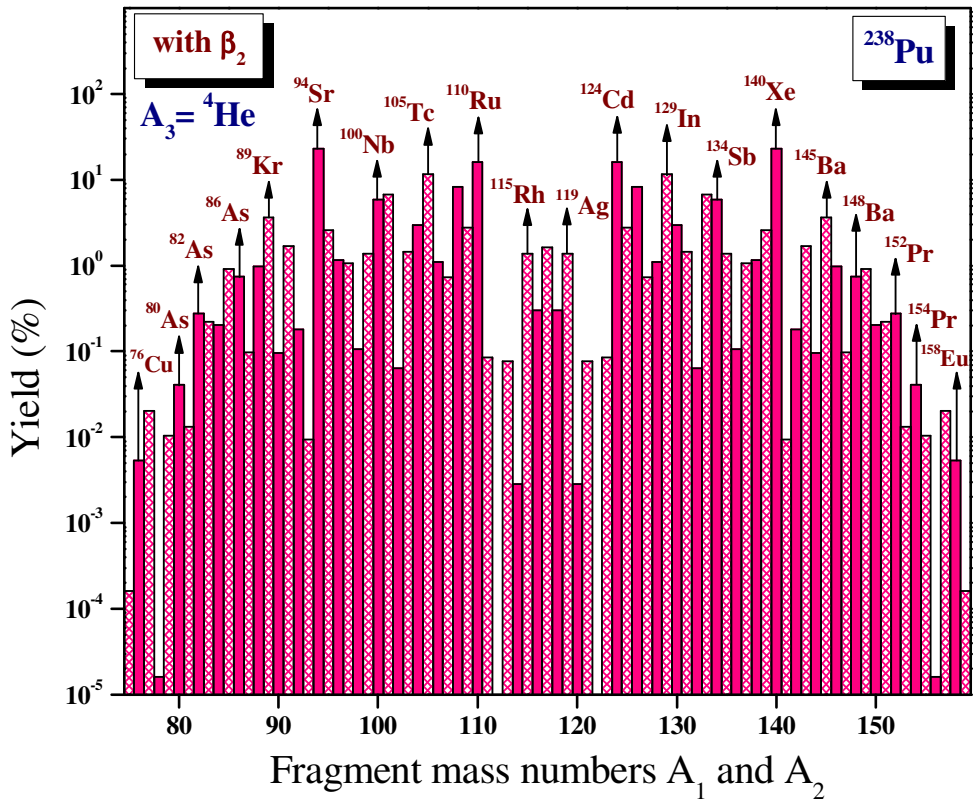

FIG 10. (Color online) The calculated yields for the charge minimized third fragment $^4$He with the inclusion of quadrupole deformation $\beta_2$ plotted as a function of mass numbers $A_1$ and $A_2$ for $^{238}$Pu isotopes. The fragment combinations with higher yields are labeled.

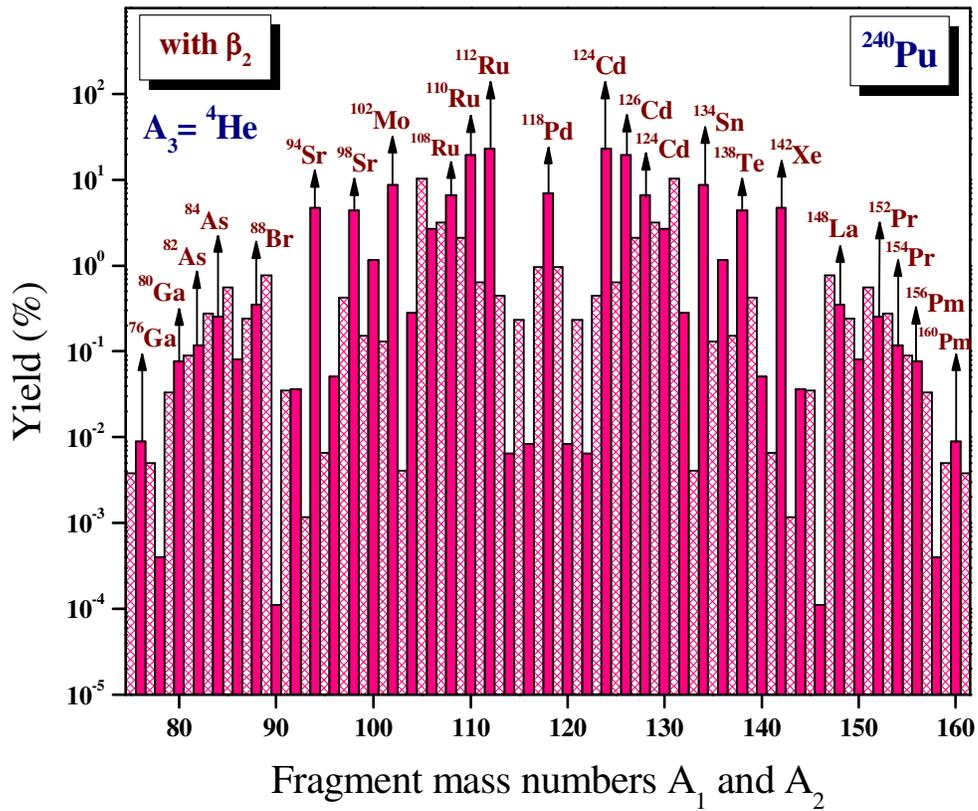

FIG 11. (Color online) The calculated yields for the charge minimized third fragment $^4$He with the inclusion of quadrupole deformation $\beta_2$ plotted as a function of mass numbers $A_1$ and $A_2$ for $^{240}$Pu isotopes. The fragment combinations with higher yields are labeled.

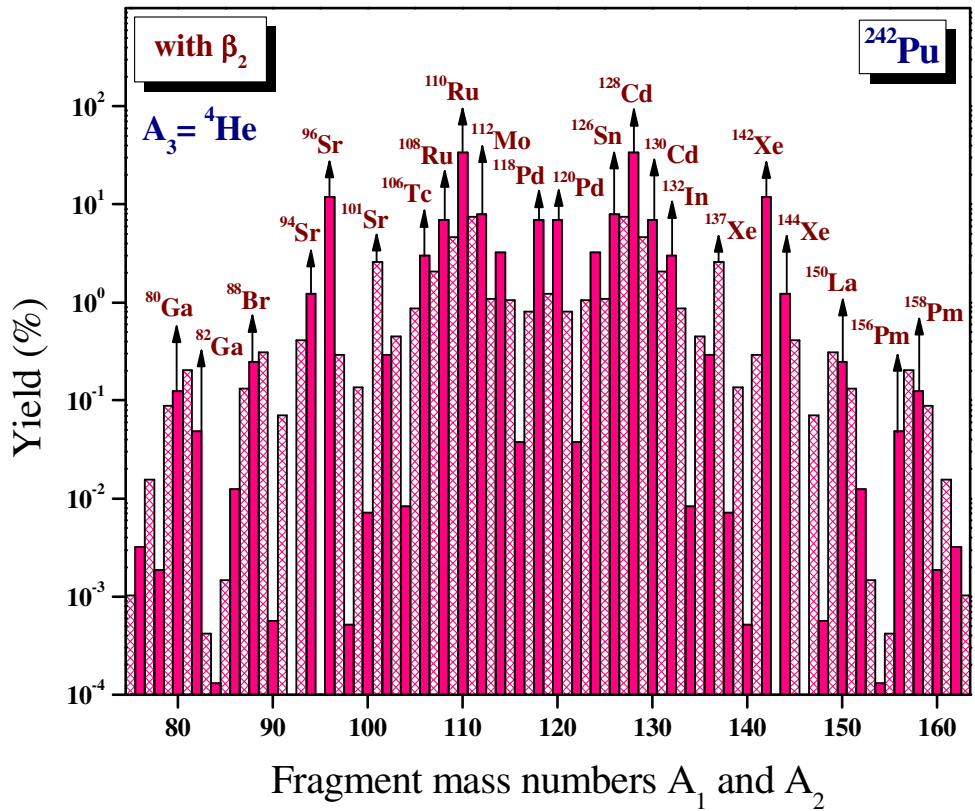

FIG 12. (Color online) The calculated yields for the charge minimized third fragment $^4$He with the inclusion of quadrupole deformation $\beta_2$ plotted as a function of mass numbers $A_1$ and $A_2$ for $^{242}$Pu isotopes. The fragment combinations with higher yields are labeled.

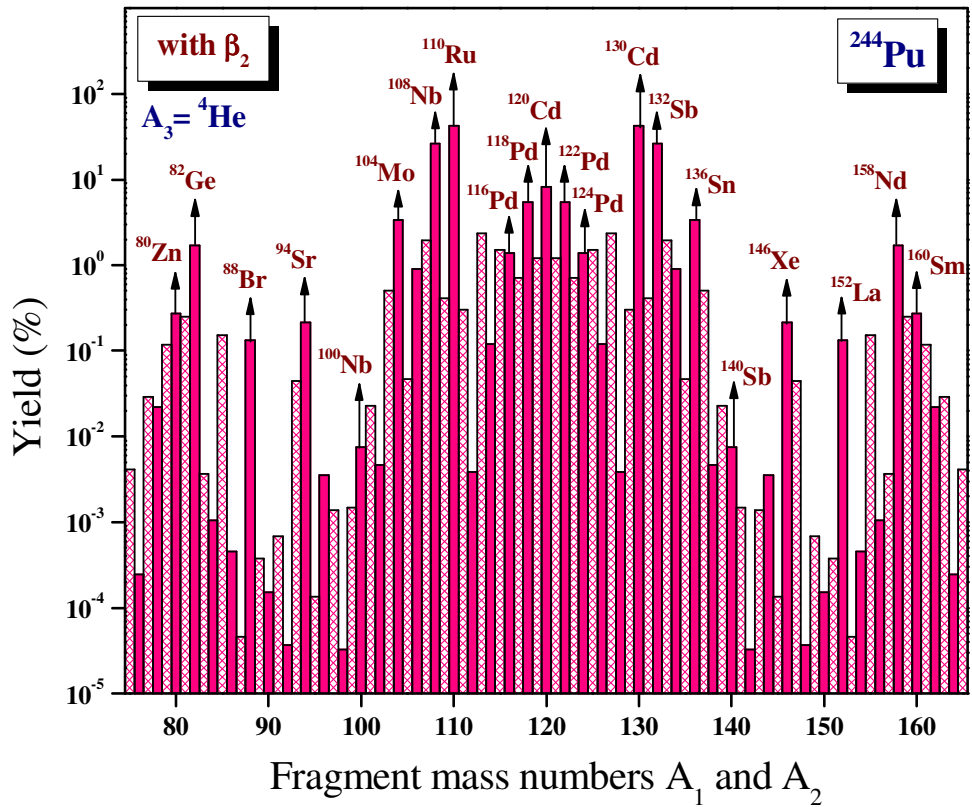

FIG 13. (Color online) The calculated yields for the charge minimized third fragment $^4$He with the inclusion of quadrupole deformation $\beta_2$ plotted as a function of mass numbers $A_1$ and $A_2$ for $^{244}$Pu isotopes. The fragment combinations with higher yields are labeled.